# In-depth characterization and analysis of simple shear flows over regularly arranged micro pillars, I. Effect of fluid inertia

Yanxing Wang[1,*], Hui Wan[2], Tie Wei[3], and Fangjun Shu[1],

[1]Department of Mechanical and Aerospace Engineering, New Mexico State University, Las Cruces, NM 88011, USA

[2]Department of Mechanical and Aerospace Engineering, University of Colorado, Colorado Springs, CO 80918, USA

[3]Department of Mechanical Engineering, New Mexico Institute of Mining and Technology, Socorro, NM 87801, USA

Through high-fidelity numerical simulation, the simple shear flow over regularly arranged micro pillars has been investigated. The essential issues to be addressed include the characteristics of a simple shear flow over quadrilateral array of micro pillars, the effect of fluid inertia on the basic flow pattern, and the decomposition of the complex surface friction. The results show that the flow is characterized by a series of microscale recirculating eddies in the gaps between the streamwise neighboring pillars. The recirculation of the micro eddies and the oscillation of the overhead flow climbing over the pillar tips create a local flow advection. At smaller Reynolds number, the fluid inertia is weak and the flow patterns are symmetrical about the pillar center. When the Reynolds number is sufficiently large, the fluid inertia takes effect and breaks the symmetrical patterns. The overhead flow tilts downward, forming a spiral long-range advection between the fluid flow above pillar array and the flow in the spaces among micro pillars. The local advection and long-range advection constitute the transport mechanism in wall-normal direction. On micro-structured walls, the total friction includes the reaction forces of micro pillars due to flow shear and flow pressure at pillar surfaces and the reaction force of bottom plane due to flow shear on bottom surface. For larger Reynolds numbers, fluid inertia prevents the fluid from flowing along the curved surface of micro pillars and reduces the equivalent shear stress of the pillar reaction force due to flow shear. At the same time, the fluid inertia makes the overhead flow impact the windward side of micro pillars more strongly and therefore increases the equivalent shear stress of the pillar reaction force due to flow pressure.

## I. INTRODUCTION

Enabled by rapid advances in micro fabrication technologies over recent decades, researchers have begun to mimic natural surfaces. The new surfaces present features on multiple scales, from the nanometer range up to the millimeter range, to optimize functional properties, such as self-cleaning, anti-adhesion, and low friction [1,2]. It has been shown that surface topologies, which are also referred to as surface roughness, can significantly alter the surface friction and heat and mass transfer on a solid wall [1,2]. Hence, they have been broadly used in microfluidic systems as a means of flow control and thermal management. Typical examples include microscale heat exchangers [3,4], chaotic mixing in microchannels [7,8], and microfluidic platform for cellular behavior study [5,6]. At such small length scales, the flow characteristics cover a very wide range of regimes, from creeping flow at very low Reynolds numbers in DNA and protein analysis devices [9], to laminar flow at low and medium Reynolds numbers in bioparticle separation systems [10], and to turbulent flow at high Reynolds numbers in the cooling systems of high-energy electronics [11]. In contrast to the booming development and application of microfluidics and surface engineering, the fundamental physics of the momentum, heat and mass exchange between the bulk flow and the structured surfaces, which are closely related to those applications, has not received sufficient attention.

Fluid dynamics on micro-structured or rough surfaces have long been one of the most important topics in fluid mechanics research. However, most of the research has been focusing on turbulent flows, and some important progress has been made [12-16]. In the laminar regime, however, the flow characteristics and the transport mechanisms for momentum, heat, and mass are still not well understood. Although many non-intrusive velocimetry methods have been developed [17,18], it remains a significant challenge to measure 3D flow characteristics with high spatiotemporal resolution at micrometer and nanometer scale. Among different topics, the characterization of surface friction and spatial variation of flow shear rate is essential and important in many applications, such as superhydrophobic drag reduction on nano- or micro-structured surfaces [19,20], in vitro microfluidic platform for flow-mediated cellular mechanotransduction analysis [21,22], and hierarchical structured antifouling surfaces [23-25]. To find out the underlying physics, a considerable amount of effort has been devoted to the fundamental research, and some experimental measurements and numerical simulations were carried out [17,26,27]. For example, Gundat et al. [26] measured the pressure drop and flow resistance in the microchannels with integrated micropillars using a pressure transducer. Evans et al. [17] used a high-speed digital holographic microscopy in combination with a correlation based de-noising algorithm to measure the wall stress and 3D flow velocity over micro-pillar array. Ichikawa et al. [27] used an astigmatism PTV to measure the 3D flow velocity and wall shear stress on a pillar-arrayed surface, and compared the results with numerical simulations. However, the results obtained in these studies are still far from enough to discover the complex mechanisms. The momentum transport from the upper flow to the structured surface is directly associated with the detailed flow evolution in the spaces among the surface structures, and is determined by the geometries and arrangement of surface structures. An exploration based on the analysis of detailed flow characteristics combined with the effects of surface topologies and Reynolds number is the key to revealing the physics.

---

*yxwang@nmsu.edu

In the research on turbulent boundary layers, it has been found that highly-regular rough surfaces can exhibit some special behaviors not common to irregular rough surfaces, such as shielding [28] and riblet effects [29]. Compared with the irregular distribution of many natural surface structures, the regular distribution provides more feasibility for controlling the flow behaviors and improving the functional properties for different applications. Due to the simple geometry and easy fabrication, microscale pillars have been broadly used as the surface structures [19-22]. Therefore, a solid surface embedded with regularly arranged microscale pillar array has been one of the standard models of micro-structured surfaces in microfluidic devices in industrial applications and scientific research.

In this paper, we develop a high-fidelity numerical model based on the lattice Boltzmann method, and extensively investigate the microscopic fluid dynamics induced by regularly arranged micro pillars. As the first part of a series of research on the transport phenomena on micro-structured surfaces, this paper has two principal objectives. The first is to characterize the flow behaviors over regularly arranged micro pillars from creeping flow (Re ~ 0.01) to laminar flow regime (Re ~ $10^2$), and the second is to identify the effect of fluid inertia on flow characteristics and momentum transfer associated with micro pillars. This paper is organized as follows. The physical model is presented in Section II. The numerical methods are described in Section III. The results are analyzed in Section IV, followed by conclusions in Section V.

## II. PHYSICAL MODEL

As shown in Fig. 1, we model an incompressible simple shear flow confined by two infinitely large parallel planes, with structured thin pillars with round tips lining the bottom plane. The finite distance between the planes prevents the development of larger-scale flow characteristics, and facilitates the identification of local flow disturbances induced by the micro pillars. The coordinates are defined as streamwise ($x$), spanwise ($y$) and wall-normal ($z$), with corresponding velocity components as $u_x$, $u_y$ and $u_z$, respectively. The distance between the two planes is denoted by $H$. The bottom plane is fixed and the top plane moves at a constant velocity $U_0$ to produce flow shear. The length and diameter of the micro pillars are denoted by $l_p$ and $D_p$, respectively. The micro pillars are arranged in a quadrilateral. The streamwise and spanwise distances between neighboring pillar centerlines are denoted as $\delta_x$ and $\delta_y$, respectively. The shear rate of the bulk flow can be roughly estimated as

$$S = U_0/H. \tag{1}$$

The Reynolds number based on the bulk flow shear rate $S$ and the pillar length is

$$Re = Sl_p^2/v, \qquad (2)$$

where $v$ is the kinematic viscosity of the fluid. In the following analysis, the pillar length ($l_p$) and equivalent velocity at pillar tips ($U_t = Sl_p$) are used as the characteristic length and velocity to normalize the spatial coordinates and flow velocity, respectively,

$$\tilde{x} = x/l_p, \qquad (3)$$

$$\tilde{\mathbf{u}} = \mathbf{u}/U_t, \qquad (4)$$

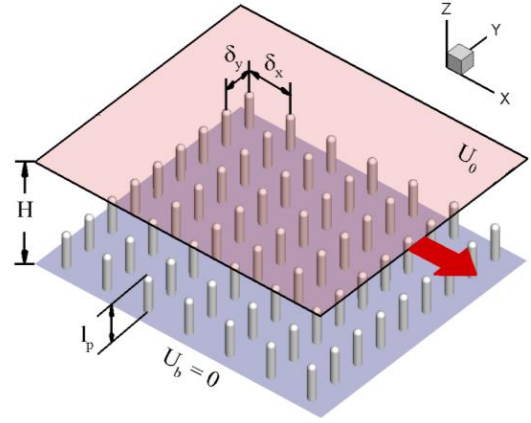

FIG. 1. Physical model of a simple shear flow with micro-pillar array on bottom plane.

In the normalized form, the pillar length is $\tilde{l}_p = 1$ and the equivalent velocity at pillar tips is $\tilde{U}_t = 1$. A complete description of the problem includes the Reynolds number ($Re$), the pillar aspect ratio ($\tilde{l}_p/\tilde{D}_p$), the distance between two planes, $\tilde{H}$, and the streamwise and spanwise distances between neighboring pillars, $\tilde{\delta}_x$ and $\tilde{\delta}_y$. In this study, $\tilde{l}_p/\tilde{D}_p$ is fixed at 5, $\tilde{H}$ is fixed at 3, and the ratios of pillar distance to diameter in streamwise and spanwise directions ($\tilde{\delta}_x/\tilde{D}_p$ and $\tilde{\delta}_y/\tilde{D}_p$) are fixed at 4. The creeping flows at $Re \ll 1$ have been well addressed [30,31]. The flow with $Re$ from $O(10^{-1})$ to $O(10^2)$, which is typical in microfluidic systems, are the focus of this paper.

## III. NUMERICAL METHODS

In this study, we used the lattice-Boltzmann method (LBM) to simulate the continuum level incompressible fluid flow over micro-pillars. LBM is well suited to the present problem because it is highly parallelizable and highly capable in dealing with complex geometries. The dependent variable is the particle distribution function $\mathbf{f}(\mathbf{x},t)$, which quantifies the probability of finding an ensemble of molecules at position $\mathbf{x}$ with velocity $\mathbf{e}$ at time $t$ [32-34]. In three dimensions, the velocity vector $\mathbf{e}$ can be discretized into 15, 19 or 27 components (referred to as D3Q15, D3Q19 and D3Q27) [32]. Here we applied the D3Q15 approach largely to minimize computational load, with the recognition that the flow Reynolds number is relatively low. To overcome the numerical instability, a multi-relaxation-time (MRT) model was utilized in the present work [35-37]. The basic idea is to take the advection in momentum space and finish the flux in velocity space. The MRT lattice Boltzmann equation reads:

$$\mathbf{f}(\mathbf{x}+\mathbf{e}\delta_t, t+\delta_t) - \mathbf{f}(\mathbf{x},t) = -\mathbf{M}^{-1}\hat{\mathbf{S}}\left(\hat{\mathbf{f}}(\mathbf{x},t) - \hat{\mathbf{f}}^{eq}(\mathbf{x},t)\right) \qquad (5)$$

where $\mathbf{M}$ is the transformation matrix, that is, $\hat{\mathbf{f}} = \mathbf{M}\mathbf{f}$, and $\hat{\mathbf{f}}^{eq}$ is the equilibrium value of the distribution function ($\hat{\mathbf{f}}$). The transformation matrix $\mathbf{M}$ can be constructed via the Gram-Schmidt orthogonalization procedure, which is given as

$$\mathbf{M} = \left(|\rho\rangle, |e\rangle, |\varepsilon\rangle, |j_x\rangle, |q_x\rangle, |j_y\rangle, |q_y\rangle, |j_z\rangle, |q_z\rangle, |3p_{xx}\rangle, |p_{ww}\rangle, |p_{xy}\rangle, |p_{yz}\rangle, |p_{zx}\rangle, |m_{xyz}\rangle\right)^T \quad (6)$$

The collision matrix $\hat{\mathbf{S}} = \mathbf{M}\mathbf{S}\mathbf{M}^{-1}$ in moment space is a diagonal matrix,

$$\hat{\mathbf{S}} = \text{diag}(0, s_1, s_2, 0, s_4, 0, s_4, 0, s_4, s_9, s_9, s_{11}, s_{11}, s_{11}, s_{14}) \qquad (7)$$

The right side of Eq. (5) describes the mixing or collision of molecules, that drives the flow to approach a local equilibrium particle distribution, $\hat{\mathbf{f}}^{eq}$. Macroscopic variables such as density $\rho$ and velocity $\mathbf{u}$ are calculated from the moments of the distribution functions,

$$\rho(\mathbf{x},t) = \sum_\alpha f_\alpha(\mathbf{x},t), \qquad \rho(\mathbf{x},t)\mathbf{u}(\mathbf{x},t) = \sum_\alpha f_\alpha(\mathbf{x},t)\mathbf{e}_\alpha \qquad (8)$$

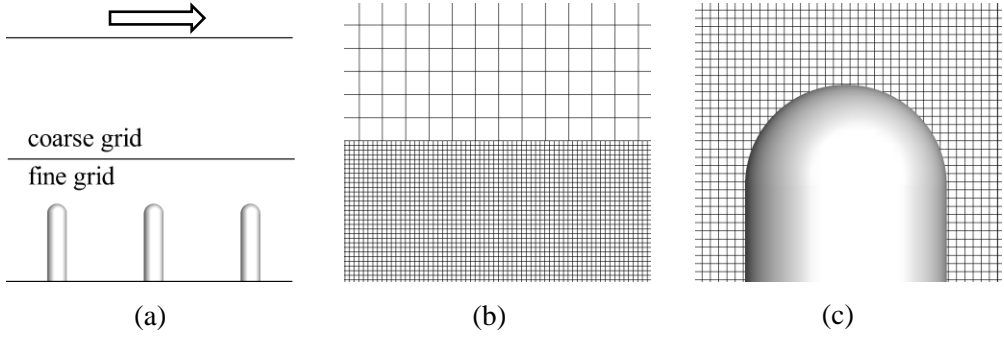

(a)          (b)          (c)

FIG. 2. Dual grid structure with a find grid around micro pillars. The ratio of coarse grid size to fine grid size is 5, and 25 grid points are used over one pillar diameter. (a) Overall grid structure, (b) transition from the fine grid to coarse grid, and (c) fine grid around a micro pillar.

In the treatment of non-slip conditions on solid boundaries, we use the scheme with 2$^{nd}$ order of accuracy advanced by Ladd [38], Bouzidi et al. [39], and Lallemand and Luo [40]. This method is based on the simple bounce-back boundary scheme and interpolations. Further details can be found in Wang et al. [34].

To reduce the computational load from an exceptionally fine uniform grid throughout the domain as required by the LBM, a dual-lattice method was utilized [34,41], with a fine grid placed in the lower region surrounding the micro pillars, as shown in Fig. 2. Within the fine grid, 25 grid points are used over the pillar diameter, which is enough to resolve the flow details considered in this study. The ratio of coarse grid size to fine grid size is $\delta x_c/\delta x_f = 5$, ensuring the smooth transition of flow quantities across the interface between the two grids. The height of the fine grid region is $\tilde{h}/\tilde{H} = 0.5$. We selected several typical cases and ran the simulations over 10 structure elements in both streamwise and spanwise directions. The flow patterns demonstrate spatial periodicities same as the distances between micro pillars in both directions. Therefore, in the large-scale systematic study, we only simulated the flow within one cuboid domain including a single pillar, and employed periodic conditions on the streamwise and spanwise surfaces. The analysis was carried out after the flow and scalar evolution entered a steady state in which all quantities remain constant over time.

The model has been extensively validated in our previous studies, and the details are described in Wang [34,42-44]. To examine the grid sensitivity of the results, simulations were conducted with three different sets of grids corresponding to different resolutions. The number of grid points over one pillar diameter is 15, 25 and 35, respectively. Figure 3 shows the profiles of the horizontally averaged streamwise velocity and vertical gradient of streamwise velocity for $\tilde{\delta}_x = \tilde{\delta}_y = 4\tilde{D}_p$ and $Re_S = 33$. Excellent convergence from the coarse to the fine grids is obtained. The maximum deviation of the quantities between the medium and the fine grids is less than 1%. Therefore, the grid with medium resolution was used in the simulations. Detailed discussion of the results will be given in Section IV.

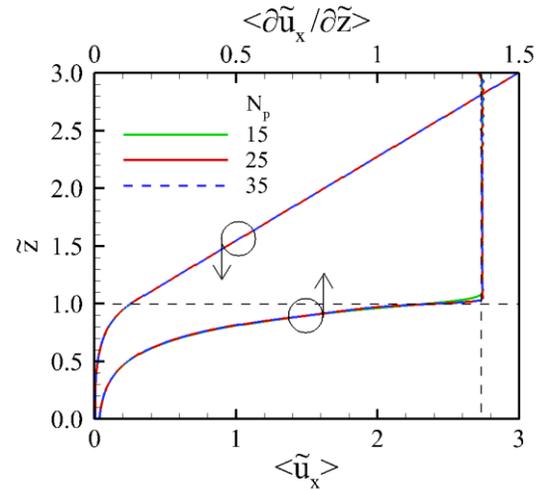

FIG. 3. Effect of grid resolution on horizontally averaged streamwise velocity $\langle \tilde{u}_x \rangle$ and vertical gradient $\langle \partial \tilde{u}_x / \partial \tilde{z} \rangle$ for $\tilde{\delta}_x = \tilde{\delta}_y = 4\tilde{D}_p$ and $Re_S = 33$. $N_p$ is the number of grid points over one pillar diameter.

## IV. RESULTS AND DISCUSSION

In this section, we first use the case of $Re = 1$ as an example to acquire the general features of the fluid flow over micropillar array, and then scrutinize the effect of fluid inertia on the flow characteristics and momentum transport details over a wider range of Reynolds number ($0.33 \leq Re \leq 100$).

### *Flow characteristics at lower Reynolds numbers*

When Reynolds number is small, such as $Re = 1$, fluid viscosity plays a dominant role in determining the flow evolution. Figure 4 gives an overview of the flow characteristics around the micro pillars at $Re = 1$. Considering the periodicities in the streamwise and spanwise directions, only the flow within a cuboid domain including a single pillar was simulated. The flow patterns containing multiple pillars shown in the figure were generated by concatenating the patterns of one pillar. In the steady state, the streamlines coincide with the trajectories of fluid particles. As shown in Fig. 4(a), the finite thickness and height of micro pillars forms a rectangular gap between each pair of the streamwise neighboring pillars, in which the velocity magnitude is reduced. The flow above the gaps drives a clockwise recirculating eddy in each gap. Basically, the size of the eddies is determined by the geometry and arrangement of micro pillars and the Reynolds number. Induced by these eddies, the flow surrounding the pillars takes an upward motion on the leeward side and a downward motion on the windward side of the pillars as it travels downstream.

Figure 4(b) shows the patterns of streamwise velocity $\tilde{u}_x$ in the flow. From the 3D iso-surfaces around the pillars and the 2D iso-contours on the streamwise and spanwise planes, it can be seen that the magnitude of streamwise velocity is significantly reduced in the spaces among the pillars. Among the pillars the iso-surface curves downward apparently, indicating the downward transport of the streamwise component of fluid momentum. The right panel in Fig. 4(b) shows the distribution of $\tilde{u}_x$ along the vertical lines at four typical positions. Line 1 coincides with the central axis of the micro pillar, Lines 2 and 3 are at the middle of the streamwise and spanwise edges of the quadrilateral on which the pillars are embedded, and Line 4 is at the center of the quadrilateral. For comparison purpose, the straight line corresponding to the profile of $\tilde{u}_x$ of the flow over a smooth wall is also shown in the figure. Beyond a certain distance above the pillar tips, $\tilde{u}_x$ changes linearly with $\tilde{z}$, and the derivative ($\partial \tilde{u}_x / \partial \tilde{z}$) is the same at four positions. Along the vertical line through pillar center (Line 1), $\tilde{u}_x$ decreases sharply to 0 towards the pillar tip. Along the lines among pillars (Lines 2, 3 and 4), $\tilde{u}_x$ decreases smoothly from a small value at the height of pillar tips to 0 at the bottom plane. This curve variation suggests that $\tilde{u}_x$ is significantly reduced below the pillar tips. For this flow, the curves between the streamwise row of pillars (Lines 3 and 4) are close to each other because of the small streamwise pillar distance $\delta_x$. Due to the shielding effect of streamwise aligned pillar rows, the variation of $\tilde{u}_x$ along the line between the streamwise neighboring pillars (Line 2) deviates apparently from the others, and the velocity magnitude is smaller in the lower region.

Fig. 4(c) shows the patterns of wall-normal velocity ($\tilde{u}_z$) induced by micro pillars. When the overhead flow climbs over the pillar tips, the upward and downward motions create a region with positive $\tilde{u}_z$ on the windward side and a region with negative $\tilde{u}_z$ on the leeward side of the hemispherical tips of micro pillars. Within the gaps between the streamwise neighboring pillars, $\tilde{u}_z$ is positive on the leeward side and is negative on the windward side, due to the clockwise recirculating eddy driven by the overhead flow (Fig. 4(a)). The magnitude of $\tilde{u}_z$ is smaller in the gaps, compared to that around the pillar tips, as shown in the Figure.

The right panel of Fig. 4(c) shows the variation of $\tilde{u}_z$ along four vertical lines surrounding a pillar. The vertical lines are symmetrically placed around a pillar, and the distance from the pillar centerline is $0.75\tilde{D}_p$. The variation of $\tilde{u}_z$ on the windward (Line 5) and leeward (Line 6) sides are largely symmetrical due to the weak fluid inertia at small Reynolds numbers. The magnitude of $\tilde{u}_z$ around the pillar tip is larger than that in the spaces, which is consistent with the patterns of $\tilde{u}_z$ shown in the left and middle panels. On the spanwise sides of the pillar, $\tilde{u}_z$ is less disturbed, so the magnitude of $\tilde{u}_z$ along Lines 7 and 8 is smaller than that along Lines 5 and 6.

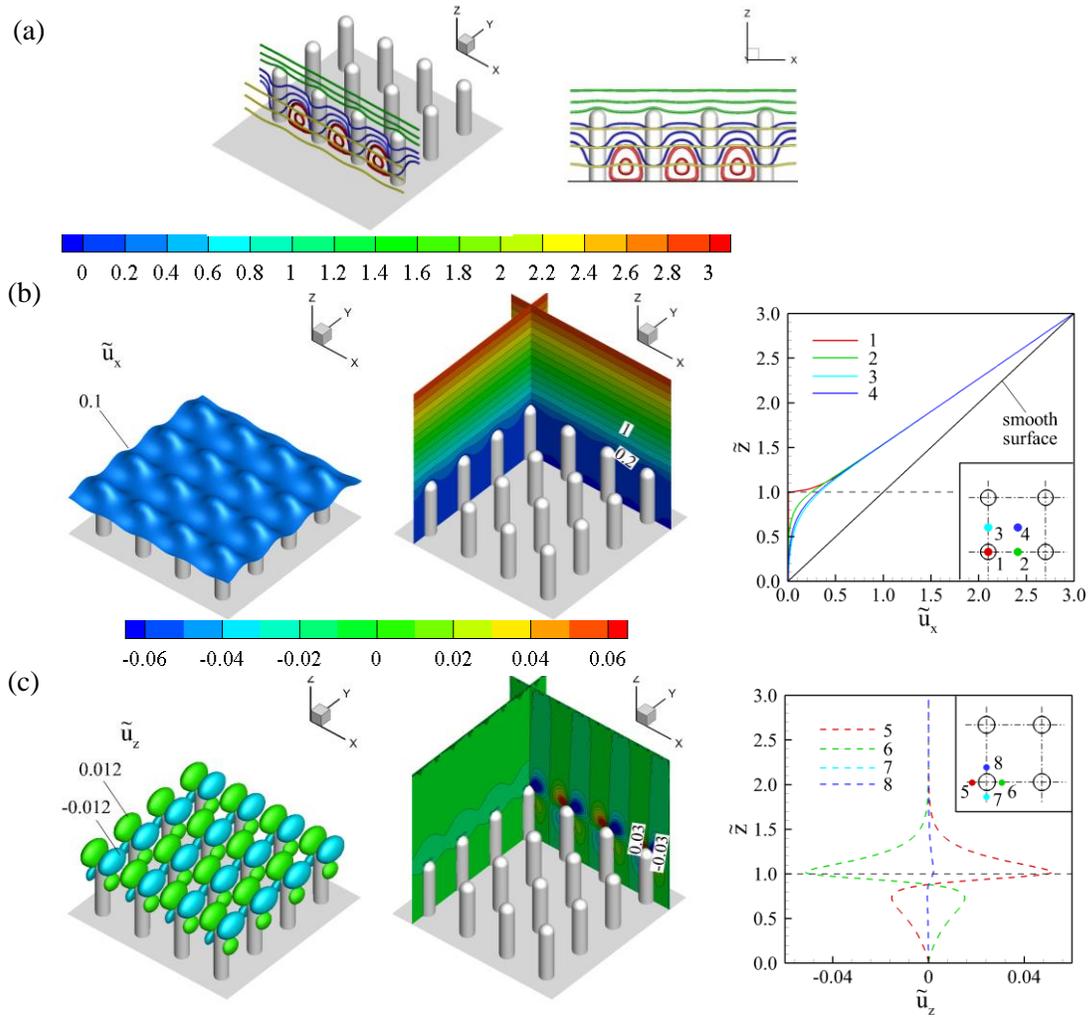

FIG. 4. Patterns of velocity characteristics for $Re = 1$. (a) Typical streamlines around micro pillars, (b) axial velocity $\tilde{u}_x$, (c) vertical velocity $\tilde{u}_z$. Left of (b) and (c): iso-surfaces, middle: iso-contours on the planes through pillar centers in axial and spanwise directions, right: distributions along vertical lines around micro pillars. Lines 1, 2, 3 and 4 are at the corner point, middle points of the edges and center point of the square, Lines 5, 6, 7 and 8 are on the edges, and the distance is $0.75\tilde{D}_p$ from the pillar center.

Flow shear rate is an important parameter in many applications, such as in vitro platform for cellular culture [21,22] and shear-induced particle margination in living organisms and medical equipment [45,46]. It also reflects the characteristics of momentum transport through fluid viscosity. For the simple shear flows considered in this study, the dominant component of the shear rate is the vertical gradient of the streamwise velocity ($\partial \tilde{u}_x/\partial \tilde{z}$). Figure 5 shows the patterns of $\partial \tilde{u}_x/\partial \tilde{z}$ around the micro pillars at $Re = 1$. As shown by the 3D iso-surfaces around the pillars and the 2D iso-contours on the streamwise and spanwise planes in Fig. 5(a), the micro pillars disturb the fluid flow in the lower region, and make the fluid trajectories suddenly change direction around the pillar tips. This sudden change in flow direction causes an increase in the magnitude of $\partial \tilde{u}_x/\partial \tilde{z}$. The regions with increased $\partial \tilde{u}_x/\partial \tilde{z}$ around pillar tips extend in the streamwise direction and connect with each other, forming a curved tubular shape above each streamwise row of pillars. The right panel shows the variation of $\partial \tilde{u}_x/\partial \tilde{z}$ along the same four vertical lines as those in

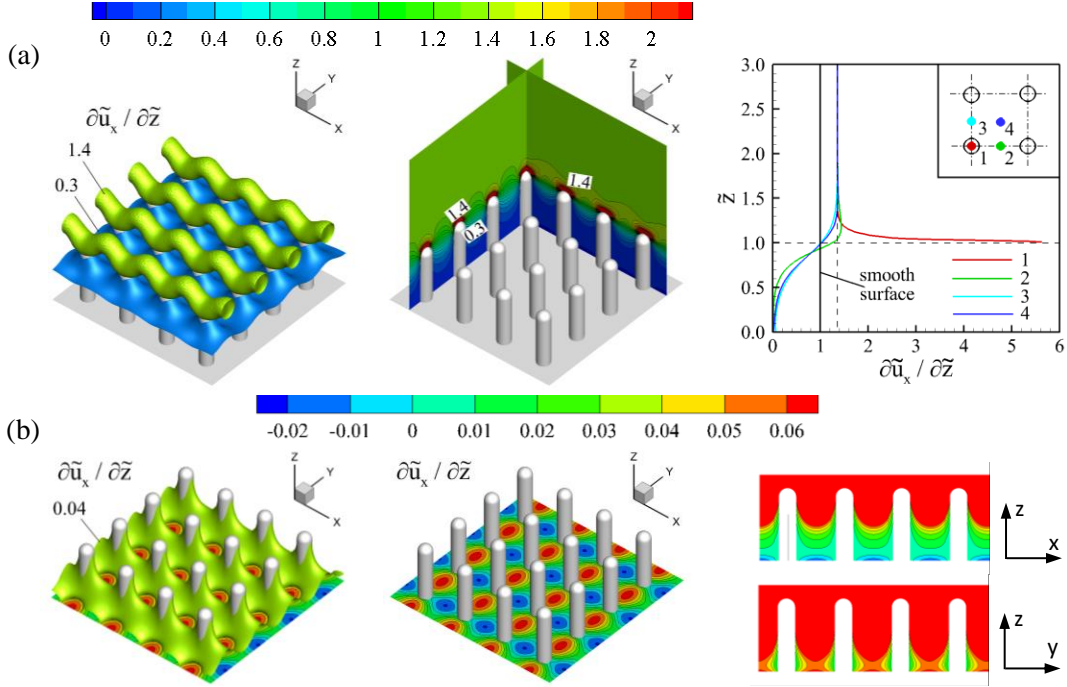

FIG. 5. Patterns of vertical gradient of axial velocity ($\partial \tilde{u}_x/\partial \tilde{z}$) for $Re = 1$. (a) Entire regime, (b) in the spaces among micro pillars.

Fig. 4(b). On a smooth wall, $\partial \tilde{u}_x/\partial \tilde{z}$ is equal to 1 throughout the flow. In the presence of pillar array, in the region beyond a certain distance above the pillar tips, $\partial \tilde{u}_x/\partial \tilde{z}$ is constant, corresponding to the linear variation of $\tilde{u}_x$ in vertical direction, as shown in Fig. 4(b). The value of $\partial \tilde{u}_x/\partial \tilde{z}$ is larger than that over a smooth surface. Along the centerline of the pillar (Line 1), $\partial \tilde{u}_x/\partial \tilde{z}$ increases sharply towards the pillar tips, corresponding to the sharp decrease of $\tilde{u}_x$ in Fig. 4(b). Along the vertical lines between the streamwise row of pillars (Lines 3 and 4), $\partial \tilde{u}_x/\partial \tilde{z}$ decreases monotonically from the constant above the pillar array to a small value at the bottom plane. Along the line between streamwise neighboring pillars (Line 2), $\partial \tilde{u}_x/\partial \tilde{z}$ first increases from the constant in the upper flow towards the height of pillar tip, and then decreases monotonically to a small value at the bottom plane. This slight increase in the region roughly above the pillar height is caused by the streamwise advection of the fluid with larger $\partial \tilde{u}_x/\partial \tilde{z}$ around the pillar tips.

Figure 5(b) shows the patterns of $\partial \tilde{u}_x/\partial \tilde{z}$ below the pillar tips. Since the streamwise velocity is significantly reduced in the spaces among micro pillars, the magnitude of $\partial \tilde{u}_x/\partial \tilde{z}$ is correspondingly smaller. The distribution of $\partial \tilde{u}_x/\partial \tilde{z}$ at the bottom surface among micro pillars is worth noting. The staggered patches of positive and negative $\partial \tilde{u}_x/\partial \tilde{z}$ are distributed at the bottom surface. Between the streamwise neighboring pillars, $\partial \tilde{u}_x/\partial \tilde{z}$ is negative. This is caused by the clockwise recirculating eddies in the gaps between the streamwise neighboring pillars, where the fluid flows in the inverse direction on the bottom surface. Between the spanwise neighboring pillars, $\partial \tilde{u}_x/\partial \tilde{z}$ is positive and the magnitude is larger. This is because that when the surrounding fluid flow passes around the pillars, the flow goes downward along the windward side of the pillars and upward along the leeward side, causing a direct impact of fluid on the bottom surface.

Figure 6 shows the flow shear rate ($|\partial \tilde{u}_s/\partial n|$, where $\tilde{u}_s$ is the tangential velocity and $n$ is the unit the normal direction) and the relative pressure ($\tilde{p}$) at pillar surface. The pressure is normalized as

$$\tilde{p} \equiv (p - p_0)/\rho U_t^2. \qquad (9)$$

where $p_0$ is the pressure at the surface of pillar root. As shown in the figure, larger $|\partial \tilde{u}_s/\partial n|$ and larger $\tilde{p}$

mainly appear at pillar tips due to the direct interaction with the overhead flow. $|\partial \tilde{u}_s/\partial n|$ reaches its maximum at the top of pillar tips. $\tilde{p}$ reach its maximum and minimum on the windward and leeward sides of the pillar, respectively. The shear stress and pressure at pillar surface provide the important components of the friction on the structured surfaces.

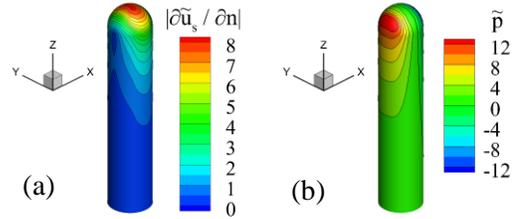

FIG. 6. Distribution of flow shear rate and pressure at the surface of a micro pillar for $Re = 1$.

At lower Reynolds numbers, such as $Re = 1$, fluid inertia is weak and the disturbance of pillar array to the ambient fluid is mainly transmitted through fluid viscosity. In this study, the geometry and arrangement of micro pillars are symmetrical in both streamwise and spanwise directions. Within the gaps between the streamwise neighboring pillars, the influence from the upstream (left) pillar is roughly the same as that from the downstream (right) pillar, which makes the flow evolution demonstrate some symmetrical features between the windward and leeward sides of the pillars about the pillar center, as shown in Fig. 4-6. As the Reynolds number increases, the effect of fluid inertia becomes stronger, which tends to maintain the fluid movement at each point and therefore breaks the symmetry of flow patterns caused by the symmetrical geometry and arrangement of micro pillars at lower Reynolds numbers.

### *Effect of fluid inertia on flow characteristics*

In order to identify the effect of fluid inertia at larger Reynolds numbers, the fluid flows at $Re = 33$ and 100 are analyzed and compared with that at $Re = 1$. Figure 7 shows the typical streamlines around the micro pillars for $Re = 1$, 33 and 100. A micro eddy is generated in each gap between the streamwise neighboring pillars for all cases. The recirculating motion of the micro eddies and the continuous upward and downward motion of the flow surrounding the pillars create a local advection in the wall-normal direction and provide a mechanism for heat and mass exchange between the bulk flow and the bottom wall. An obvious outcome of the increase in Reynolds number is the increase in the size of the micro eddies. This trend is a typical feature of the flow around fixed objects, where the stronger fluid inertia at larger Reynolds numbers prevents the fluid from flowing along the curved surfaces more strongly, thereby forming a larger recirculating eddy in the leeward area. Another outcome of increasing $Re$ is that the streamlines lose the symmetrical pattern. For $Re = 33$ and 100, the increased fluid inertia tends to maintain the downward movement of the overhead fluid after climbing over the semispherical pillar tips. This causes the overhead fluid to tilt downward as a whole as it flows downstream. Due to the continuous downward push of the overhead flow over each gap, the flow surrounding the micro pillars also tilts downward and gradually approaches the bottom surface. In the area between the streamwise row of pillars, the fluid tilts upward

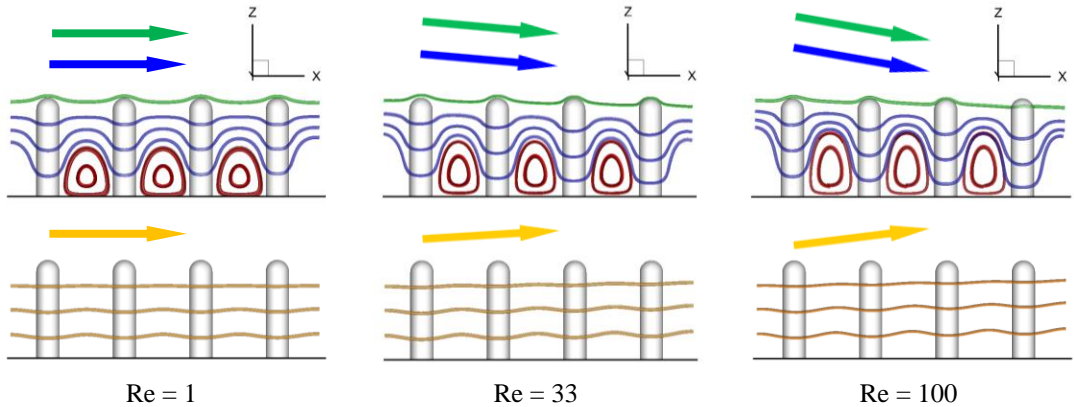

Re = 1       Re = 33       Re = 100

FIG. 7. Typical streamlines around micro pillars for $Re = 1$, 33 and 100.
Upper: around streamwise pillar row, lower: between streamwise pillar row.

when flowing downstream, as shown in the figure. It should be noted that this overall downward and upward tilt of the streamlines does not violate the spatial periodicity with a period same as the streamwise pillar distance. The spatial periodic characteristics are manifested as the periodic spatial variation of flow quantities such as velocity components. However, the streamlines are obtained by integrating the velocity components. They may not have the same periodicity.

Figure 8 shows the typical streamlines around the pillars over a longer distance for $Re = 1, 33$ and $100$. Such a long distance makes the downward tilt of the flow along the streamwise pillar rows and the upward tilt of the flow between the pillar rows more obvious. For $Re = 1$, the streamlines do not show an obvious deviation from the constant vertical level due to the weak fluid inertia. For $Re = 33$ and $100$, the increased fluid inertia not only causes the downward tilt of the flow along the streamwise pillar array, but also causes the lateral tilt from the streamwise pillar rows to the middle area between the pillar rows when the flow approaches the bottom surface. As a result of mass conservation, the flow goes upward in the area between the streamwise pillar rows and gets out of the spaces among micro pillars. In the upper region, the upward fluid flow is expected to turn back to the area above the streamwise pillar rows and then go downward. This consecutive downward movement of flow along the streamwise pillar rows and upward movement of the flow between the streamwise pillar rows produces a spiral pattern of fluid trajectories which extends in the streamwise direction. The spiral fluid trajectories create a long-range advection which might provide another mechanism of advective heat and mass transport in the wall-normal direction. The spiral streamline pattern forms a flow circulation between the upper bulk flow and the flow within the spaces among micro pillars. Comparison of the streamline patterns suggests that the downward and upward tilt becomes move obvious as the Reynolds number increases. Therefore, the long-range advection is stronger for larger Reynolds numbers. For flows at lower Re, for example $Re = 1$, each streamline only exhibits the oscillation induced by the regularly arranged micro pillars and the small eddies between the pillars, and does not show any visible vertical and lateral deviation over a long distance. This means that the long-range advection and the spiral circulation between the upper bulk flow and the lower flow in the spaces are weak at small Reynolds numbers.

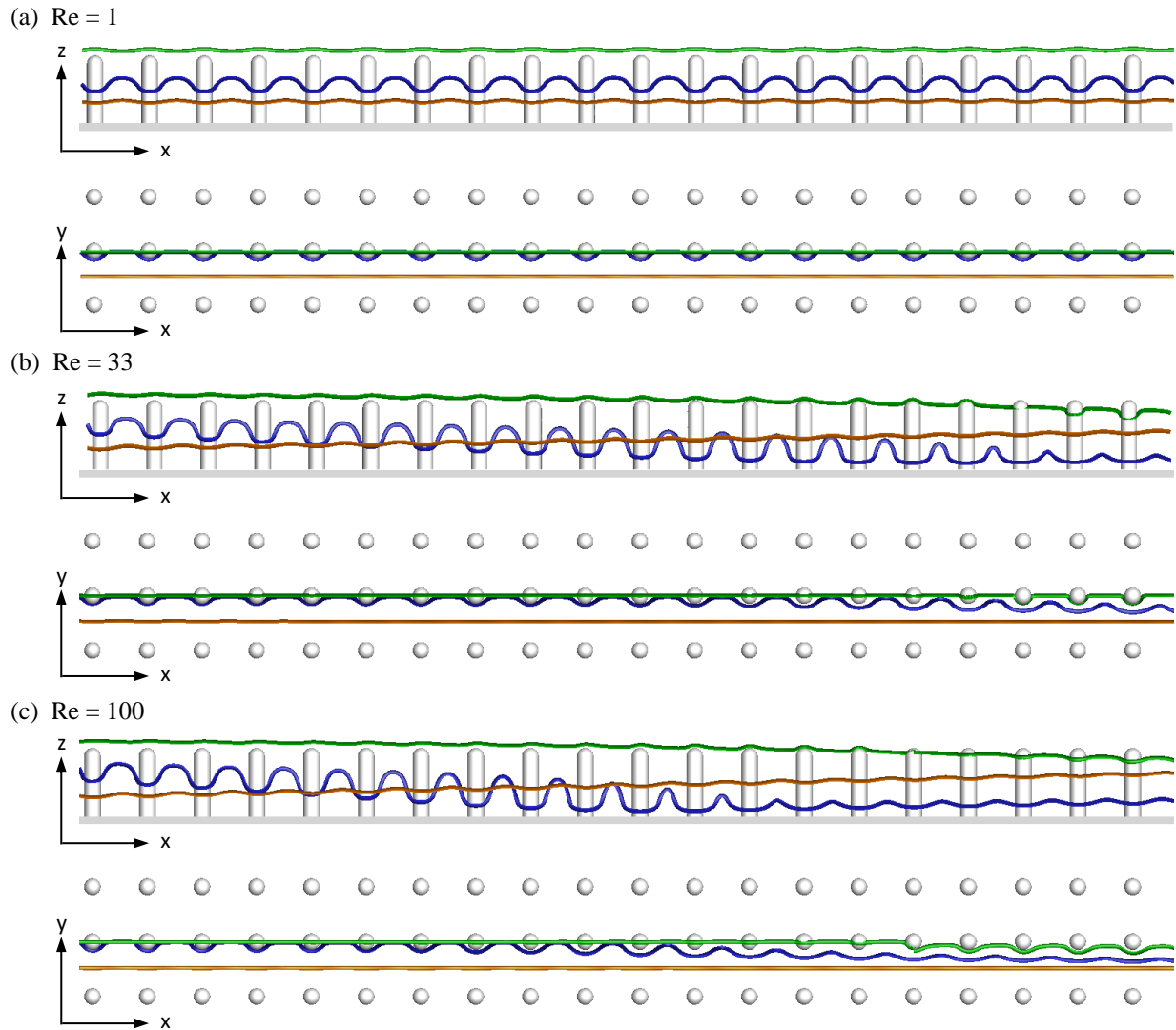

FIG. 8. Streamlines of overhead flow (green), flow surrounding micro pillars (blue), and flow between streamwise pillar array (yellow) for (a) $Re = 1$, (b) 33, and (c) 100.

To further explore the effect of fluid inertia, the velocity components are examined. Figure 9 shows the patterns of streamwise velocity $\tilde{u}_x$ for $Re = 1$, 33 and 100. The 3D iso-surfaces around the micro pillars and the 2D iso-contours on the streamwise plane through pillar centers do not show obvious difference among the three Reynolds numbers. However, on the horizontal planes at pillar height ($\tilde{z} = 1$) and at half of pillar height ($\tilde{z} = 0.5$), the 2D iso-contours of $\tilde{u}_x$ show apparent decrease in the magnitude of $\tilde{u}_x$ as $Re$ increases, especially in the area between the streamwise rows of micro pillars. In other words, the fluid flow within the spaces among micro pillars is less active in terms of streamwise velocity at larger Reynolds numbers. This seems to be contrary to the usual understanding of the effect of Reynolds number. In fact, this seemingly abnormal phenomenon is caused by the complex effect of fluid inertia on the fluid flow over surface structures. The downward tilt of overhead flow at larger $Re$ leads to a stronger impact on the windward sides of pillar tips (Figs. 7 and 8), which makes the flow turn more sharply at the surface of pillar tips and causes more loss in the streamwise momentum of the flow. Therefore, the momentum flux transferred to the fluid in the spaces among the pillars is smaller for larger Reynolds numbers, and the magnitude of streamwise velocity component is smaller correspondingly.

The fluid inertia also influences the vertical component of flow velocity. Figure 10 shows the patterns of vertical velocity $\tilde{u}_z$ for $Re = 1$, 33 and 100. For $Re = 1$, the distribution of $\tilde{u}_z$ on the windward and leeward sides of the pillars exhibits an anti-symmetrical pattern about the pillar center. The magnitude of $\tilde{u}_z$ in the spaces among micro pillars is smaller than that around the pillar tips. The anti-symmetrical pattern of $\tilde{u}_z$, especially the distribution of alternating positive and negative values on the horizontal planes, imply that the variation of $\tilde{u}_z$ contains only the oscillation caused by the regularly arranged micro pillars and the micro eddies in the spaces, that is, the fluid flow at small Reynolds numbers only has local advection. This is consistent with the observation of streamwise velocity. When $Re$ increases to 33, the downward tilt of the overhead flow due to the increased fluid inertia (Fig. 7 and 8) changes the anti-symmetrical distribution of $\tilde{u}_z$. Around the pillar tips, the downward tilt of the overhead flow weakens the movement of climbing over the pillar tips, thus reducing the magnitude of $\tilde{u}_z$ on both the windward and leeward sides of pillar tips. In the spaces among micro pillars, the downward tilt of the overhead flow strengthens the downward motion of the fluid along the windward side of the pillars, so the magnitude of $\tilde{u}_z$ is increased on the windward side. However, the magnitude of the upward velocity on the leeward side in the spaces does not increase correspondingly, but decreases significantly. This is because the long-range advection due to the increased fluid inertia modifies the spatial distribution of $\tilde{u}_z$. The fluid flow of long-range advection goes upward in the area between the streamwise rows of micro pillars, which increases the horizontal area of the region with upward velocity. As a result, the magnitude of upward velocity is reduced. On the horizontal plane at pillar height ($\tilde{z} = 1$), the area with downward velocity ($\tilde{u}_z < 0$) increases between the streamwise neighboring pillars, yet the area with upward velocity ($\tilde{u}_z > 0$) increases between the streamwise pillar rows, due to the development of long-range advection. When $Re$ increases to 100, the effect of fluid inertia becomes more prominent. The area with downward velocity further increases between the streamwise neighboring pillars on the horizontal plane at $\tilde{z} = 1$, and the discrete areas with upward velocity between the streamwise pillar rows also increase and connect with each other to form a continuous larger area. The change of vertical velocity reflects the strengthening of long-range advection with the increased Reynolds number.

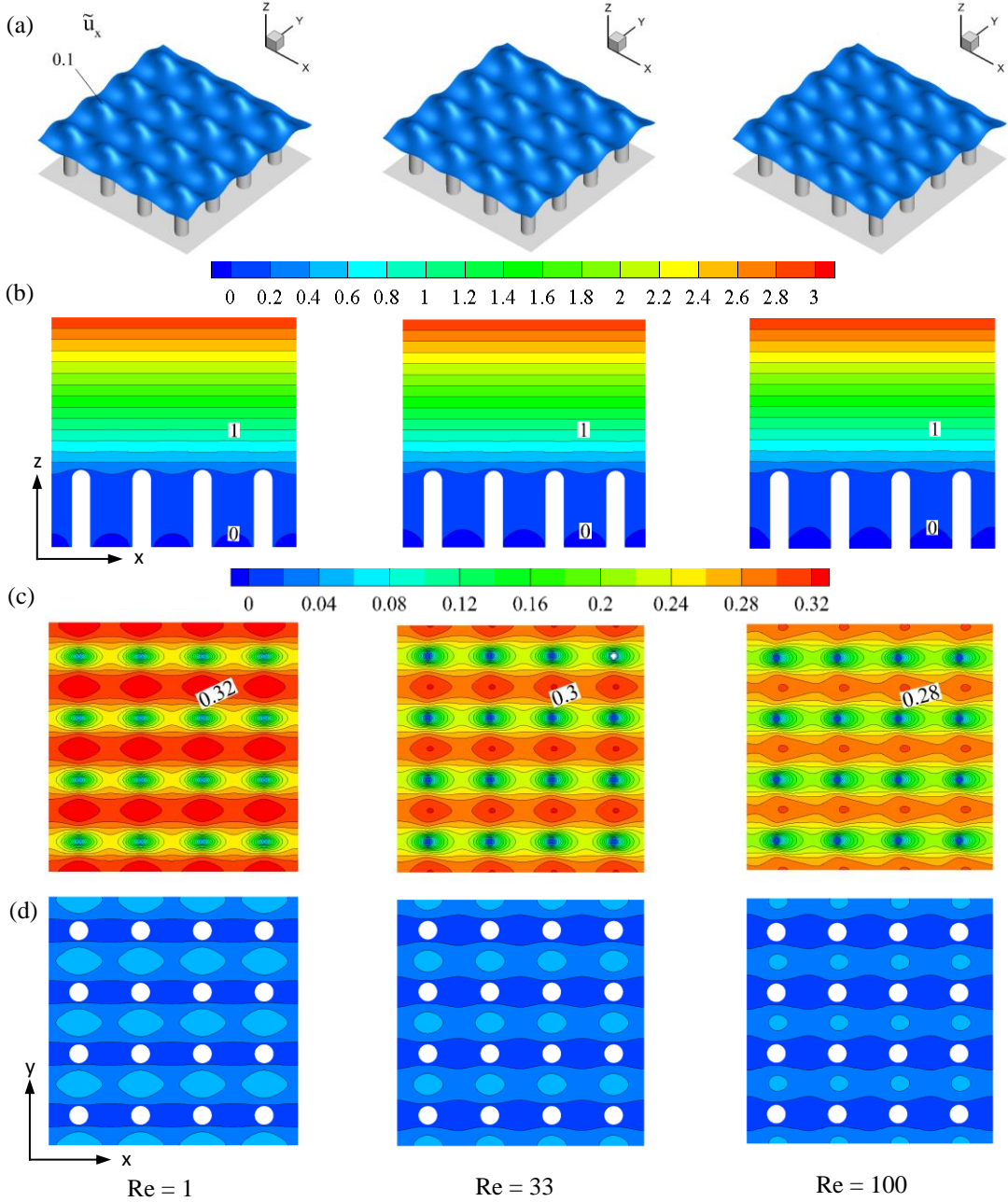

FIG. 9. Patterns of streamwise velocity $\tilde{u}_x$ for $Re = 1$, 33 and 100. Row 1 (a) iso-surfaces, (b) iso-contours on a streamwise plane through pillar centers, (c) iso-contours on a horizon planes at pillar tips ($\tilde{z} = 1$), (d) iso-contours on a horizontal plane at the midpoint of micro pillars ($\tilde{z} = 0.5$).

Figure 11 shows the vertical gradient of streamwise velocity ($\partial \tilde{u}_x / \partial \tilde{z}$) for $Re = 1$, 33 and 100. For $Re = 1$, the distribution of $\partial \tilde{u}_x / \partial \tilde{z}$ on the windward and leeward sides is symmetrical about the pillar center. When $Re$ increases to 33, the strengthened fluid inertia breaks the symmetry. The pattern of $\partial \tilde{u}_x / \partial \tilde{z}$ shifts downstream apparently, as shown by the 3D iso-surfaces around the pillars and the 2D iso-contours on the streamwise plane through pillar centerlines. At the same time, the magnitude of $\partial \tilde{u}_x / \partial \tilde{z}$ increases slightly around the pillar tips. However, due to more streamwise momentum loss caused by stronger impact of the overhead flow on pillar tips (Fig. 7 and 8), the magnitude of the streamwise velocity ($\tilde{u}_x$) is smaller in the spaces among micro pillars (Fig. 9). Correspondingly, the magnitude of $\partial \tilde{u}_x/\partial \tilde{z}$ is smaller in the spaces at larger $Re$, as shown in the figure. For the same reason, the magnitude of $\partial \tilde{u}_x/\partial \tilde{z}$ is also smaller on the bottom surface for larger $Re$. When $Re$ increases to 100, the aforementioned effect of fluid inertia becomes stronger. The magnitude of $\partial \tilde{u}_x/\partial \tilde{z}$ around pillar tips becomes larger, as shown by the 2D iso-contours on the horizontal plane at pillar height. On the other hand, the magnitude of $\partial \tilde{u}_x/\partial \tilde{z}$ in the spaces among micro pillars and on the bottom surface becomes smaller. One thing should be noted is that the downward tilt of overhead flow suppresses the expansion of the area with larger $\partial \tilde{u}_x/\partial \tilde{z}$ around the pillar tips. Therefore, the tubular structures of the 3D iso-surfaces become smaller at larger $Re$.

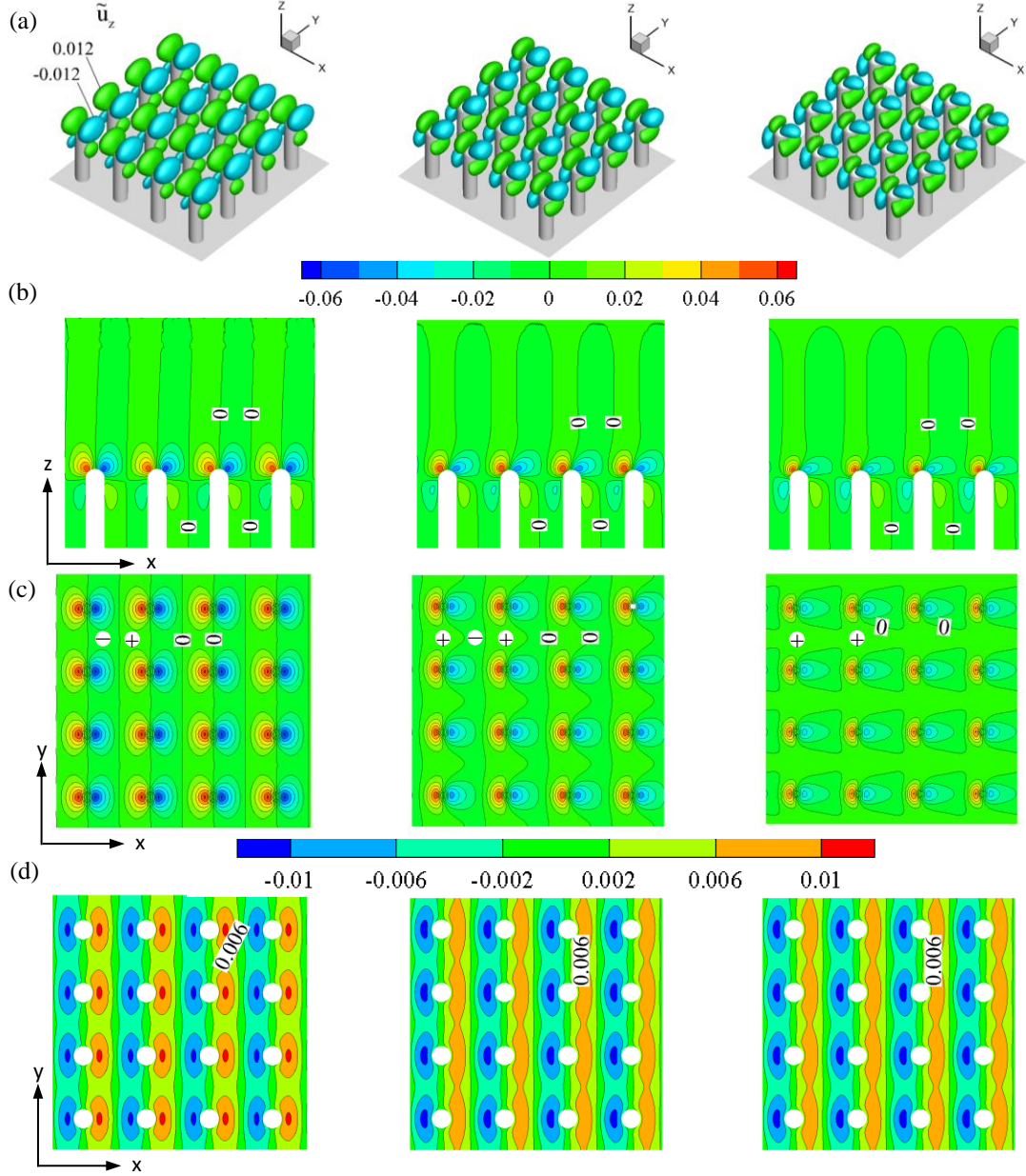

FIG. 10. Patterns of vertical velocity $\tilde{u}_z$ for $Re = 1$, 33 and 100. Row 1 (a) iso-surfaces, (b) iso-contours on a streamwise plane through pillar centers, (c) iso-contours on a horizon planes at pillar tips ($\tilde{z} = 1$), (d) iso-contours on a horizontal plane at the midpoint of micro pillars ($\tilde{z} = 0.5$).

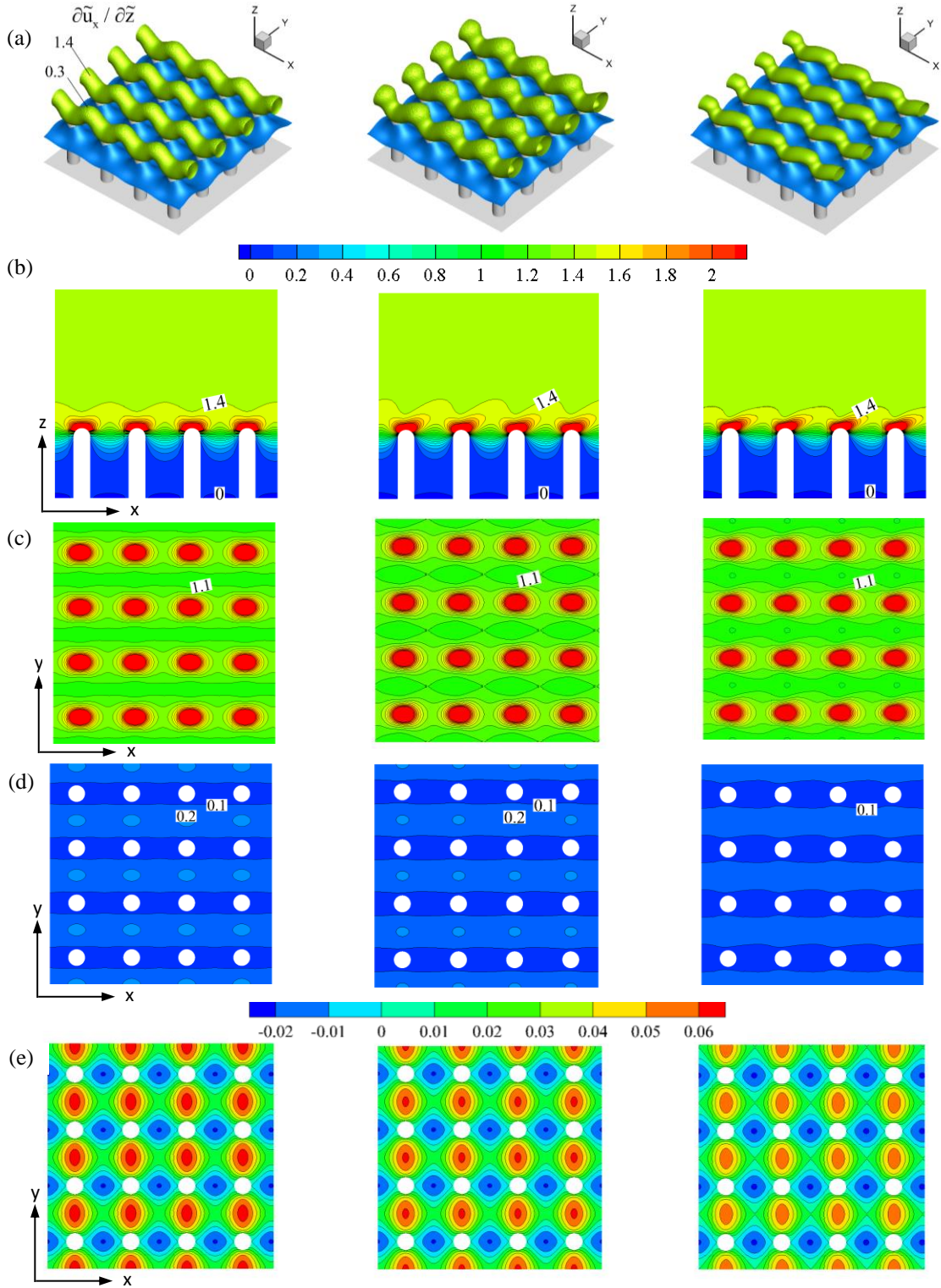

FIG. 11. Patterns of vertical gradient of streamwise velocity ($\partial \tilde{u}_x/\partial \tilde{z}$) for $Re = 1$, 33 and 100. (a) iso-surfaces, (b) iso-contours on a streamwise plane through pillar centers, (c) iso-contours on a horizon planes at pillar tips ($\tilde{z} = 1$), (d) iso-contours on a horizontal plane at the midpoint of micro pillars ($\tilde{z} = 0.5$), (e) iso-contours on the bottom plane ($\tilde{z} = 0$).

Figure 12 shows the variations of streamwise ($\tilde{u}_x$) and vertical ($\tilde{u}_z$) velocities and vertical gradient of streamwise velocity ($\partial \tilde{u}_x/\partial \tilde{z}$) along several typical vertical lines for $Re = 1$, 33 and 100. The vertical lines are labelled in the same way as that in Fig. 4. Fig. 12(a) plots $\tilde{u}_x$ along the vertical lines at the middle points of the streamwise (Line 2) and spanwise (Line 3) neighboring pillars. Due to the shielding effect of regularly arranged micro pillars, the magnitude of $\tilde{u}_x$ on Line 2 is smaller than that on Line 3. On both lines, $\tilde{u}_x$ is smaller for larger $Re$, which is consistent with the patterns shown in Fig. 9. Near the bottom surface, $\tilde{u}_x$ becomes negative on Line 2 due to the reverse flow in the micro eddy. Because of the smaller magnitude of $\tilde{u}_x$, $\partial \tilde{u}_x/\partial \tilde{z}$ is also smaller for larger $Re$ on both lines, as shown in Fig. 12(b). Near the bottom surface, $\partial \tilde{u}_x/\partial \tilde{z}$ becomes negative on Line 2 because of the micro eddy. Fig. 12(c) shows the variation of $\tilde{u}_z$ on the vertical lines on the windward (Line 5) and leeward (Line 6) sides of the pillar, respectively. Around the pillar tips, the distribution of $\tilde{u}_z$ on Lines 5 and 6 are roughly symmetrical about the pillar center, and the magnitude of $\tilde{u}_z$ is larger for smaller Reynolds numbers. Below the pillar tips, the downward motility of the overhead flow increases the magnitude of downward velocity on the windward side (Line 5) as $Re$ increases. Yet on the leeward side (Line 6), the magnitude of $\tilde{u}_z$ decreases slightly due to the development of long-range advection (see Fig. 11). Near the bottom surface, the magnitude of $\tilde{u}_z$ is larger for smaller $Re$ on both sides.

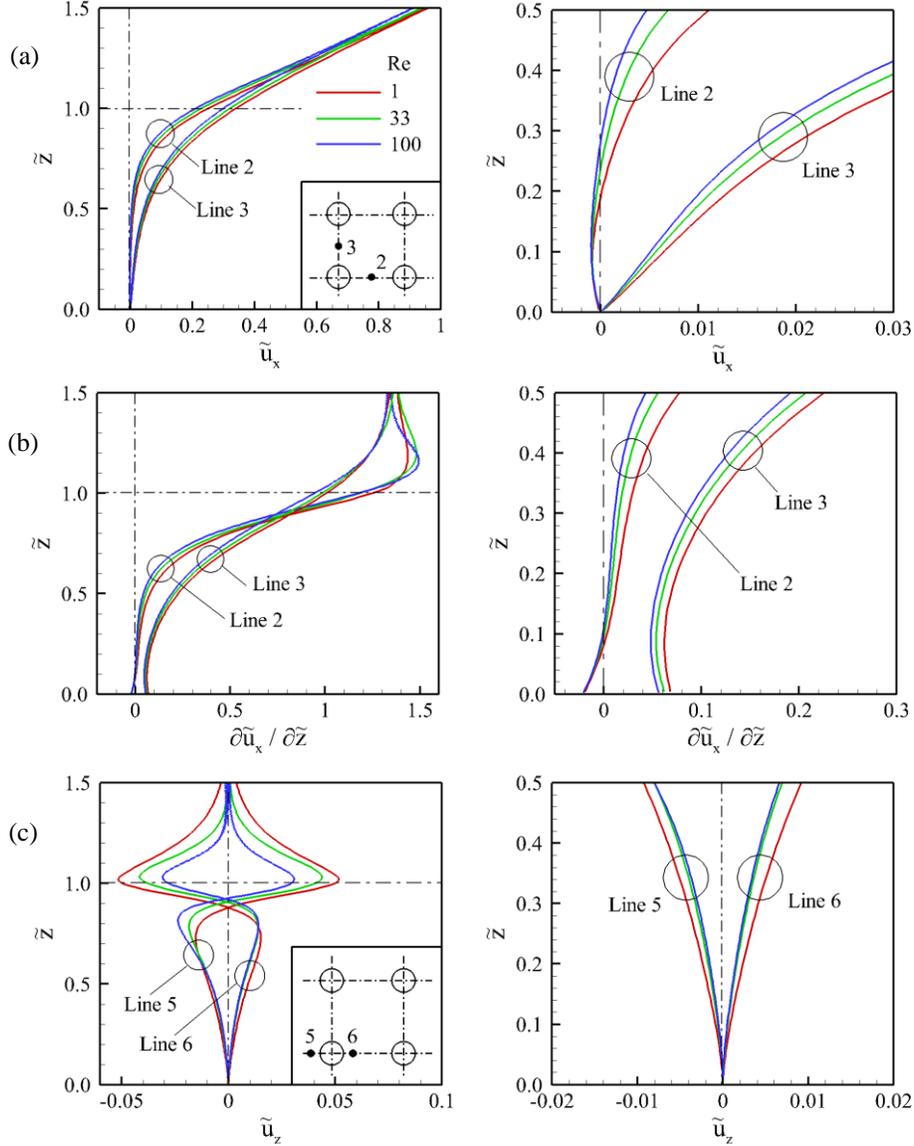

FIG. 12. Distributions of (a) streamwise velocity ($\tilde{u}_x$) and (b) vertical gradient of streamwise velocity ($\partial \tilde{u}_x/\partial \tilde{z}$) along the vertical lines at the middle points of streamwise (Line 2) and spanwise (Line 3) neighboring pillars, and (c) vertical velocity ($\tilde{u}_z$) along the vertical lines on the windward and leeward sides of micro pillars, whose distance from pillar center is $0.75\tilde{D}_p$. The right column gives the distributions in the lower region ($\tilde{z} < 0.5$).

In order to compare the overall variation of flow quantities and the momentum transport processes of different flows, we define the horizontally averaged quantities as,

$$\langle \tilde{\beta} \rangle(\tilde{z}) \equiv \frac{1}{A_f} \int_{A_f} \tilde{\beta}(\tilde{x}, \tilde{y}, \tilde{z}) d\tilde{x} d\tilde{y} \qquad (10)$$

where $\tilde{\beta}(\tilde{x}, \tilde{y}, \tilde{z})$ is the quantity of interest, and $A_f$ is the area of horizontal plane occupied by fluid. The horizontally averaged quantity ($\langle \tilde{\beta} \rangle$) is only a function of vertical coordinate $\tilde{z}$. Figure 13 shows the variation of horizontally averaged quantities for $Re = 1, 33$ and $100$. Fig. 13(a) shows the streamwise velocity $\langle \tilde{u}_x \rangle$. Above the pillar tips ($\tilde{z} > 1$), $\langle \tilde{u}_x \rangle$ changes linearly with vertical coordinate, from 3 at the top plane to a small value at the pillar tips. Below the pillar tips ($\tilde{z} < 1$), $\langle \tilde{u}_x \rangle$ decreases to zero asymptotically towards the bottom plane. Over the whole range of $\tilde{z}$, the magnitude of $\langle \tilde{u}_x \rangle$ is slightly smaller for larger $Re$, because of more momentum loss at pillar tips as analyzed in Fig. 7-9. Correspondingly, the vertical gradient of streamwise velocity ($\langle \partial \tilde{u}_x / \partial \tilde{z} \rangle$) remains constant above the pillar tips for all Reynolds numbers, and decreases to a small but nonzero value towards the bottom plane, as shown in Fig. 13(b). Near the pillar tips, $\langle \partial \tilde{u}_x / \partial \tilde{z} \rangle$ decreases sharply, corresponding to the strong momentum loss at pillar tips. Above the pillar tips, $\langle \partial \tilde{u}_x / \partial \tilde{z} \rangle$ is slight larger for larger $Re$, because the more momentum loss at pillar tips induces a larger momentum flux from the top plane. Yet below the pillar tips, the more momentum loss at pillar tips reduces the momentum transfer to the bottom surface. As a result, $\langle \partial \tilde{u}_x/\partial \tilde{z} \rangle$ is smaller for larger $Re$. Figure 13(c) shows the second derivative of streamwise velocity with respect to vertical coordinate ($\langle \partial^2 \tilde{u}_x/\partial \tilde{z}^2 \rangle$), which describes the loss rate of streamwise momentum at each vertical level due to the interaction with the micro pillars. Although the vertical flow advection also has contribution to the variation of momentum, but its fraction is relatively lower as shown in Fig. 13(d). Above the pillar tips ($\tilde{z} > 1$), $\langle \partial^2 \tilde{u}_x/\partial \tilde{z}^2 \rangle$ is close to 0 for all Reynolds numbers. Near the pillar tips, $\langle \partial^2 \tilde{u}_x/\partial \tilde{z}^2 \rangle$ increases sharply due to the strong interaction between the fluid flow and the pillar tips. Below the pillar tips ($\tilde{z} < 1$), $\langle \partial^2 \tilde{u}_x/\partial \tilde{z}^2 \rangle$ decreases asymptotically to a small value towards to the bottom surface, suggesting the momentum loss due to the interaction with the pillars decreases gradually towards the pillar roots. Around the pillar tips $\langle \partial^2 \tilde{u}_x/\partial \tilde{z}^2 \rangle$ is larger for larger $Re$, yet below the pillar tips ($\tilde{z} < 1$) $\langle \partial^2 \tilde{u}_x/\partial \tilde{z}^2 \rangle$ is smaller for larger $Re$. This variation suggests that the stronger interaction at the pillar tips at larger $Re$ reduces the momentum loss rate below the pillar tips. Fig. 13(d) shows the and advective flux of streamwise momentum in vertical direction ($\langle \tilde{u}_z \partial \tilde{u}_x/\partial \tilde{z} \rangle$). Around the pillar tips, the magnitude of $\langle \tilde{u}_z \partial \tilde{u}_x/\partial \tilde{z} \rangle$ is larger, corresponding to the stronger flow oscillation. In the spaces, the magnitude is much smaller. Compared with $\langle \partial^2 \tilde{u}_x/\partial \tilde{z}^2 \rangle$, the magnitude of $\langle \tilde{u}_z \partial \tilde{u}_x/\partial \tilde{z} \rangle$ is much smaller, which implies the vertical advection plays a minor role in the process of momentum transport.

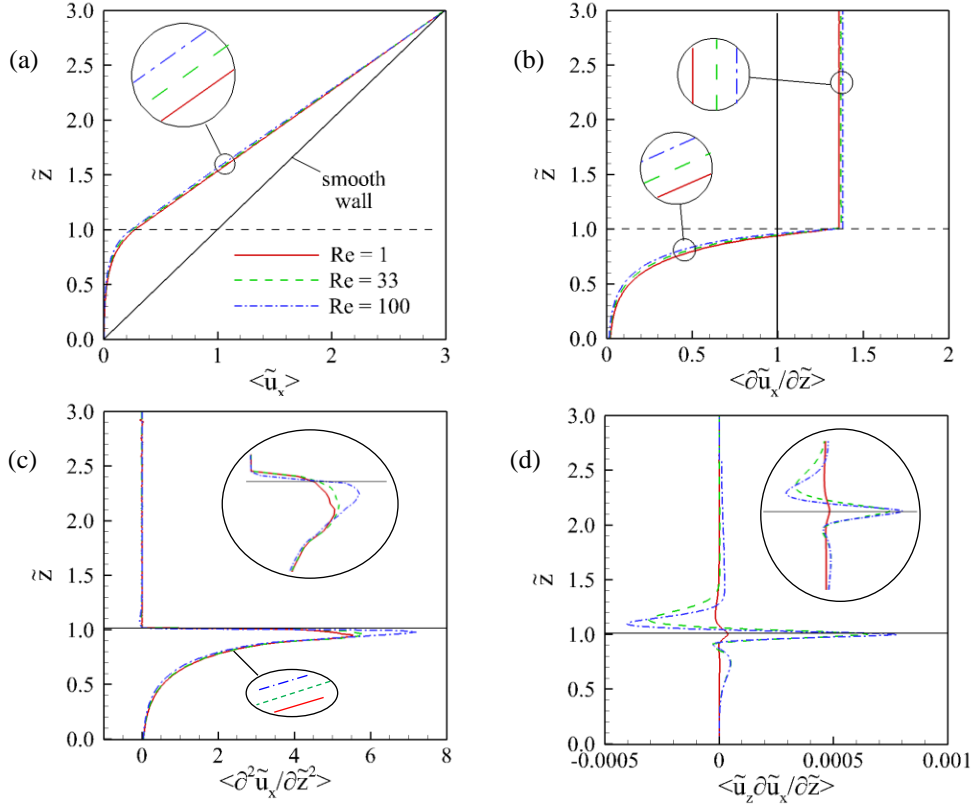

FIG. 13. Distributions of horizontally averaged quantities for $Re = 1, 33$ and $100$. (a) Axial velocity $\langle \tilde{u}_x \rangle$, (b) vertical gradient of streamwise velocity $\langle \partial \tilde{u}_x / \partial \tilde{z} \rangle$, (c) 2$^{nd}$ order gradient of streamwise velocity $\langle \partial^2 \tilde{u}_x / \partial \tilde{z}^2 \rangle$, and (d) advective flux of streamwise momentum $\langle \tilde{u}_z \partial \tilde{u}_x / \partial \tilde{z} \rangle$.

### *Decomposed surface friction on micro-structured wall*

On micro-structured surfaces, the total hydrodynamic drag ($F_{tot}$) over each structure unit consists of three components: the reaction force of a single pillar due to the flow shear stress at pillar surface ($F_{p,s}$), the reaction force of a single pillar due to the flow pressure at pillar surfaces ($F_{p,p}$), and the reaction force of the bottom plane due to the flow shear stress at bottom surface ($F_{b,s}$), that is,

$$F_{tot} = F_{p,s} + F_{p,p} + F_{b,s} \tag{11}$$

$F_{p,s}$, $F_{p,p}$ and $F_{b,s}$ are calculated by integrating the streamwise components of shear stress and pressure on pillar surface and bottom surface, respectively,

$$F_{p,s} = \int_{A_p} \left(\mu \frac{du_s}{dn}\right) s_x dA \tag{12}$$

$$F_{p,p} = -\int_{A_p} p n_x dA \tag{13}$$

$$F_{b,s} = -\int_{A_b} \left(\mu \frac{du_s}{dn}\right) s_x dA \tag{14}$$

where $A_p$ and $A_b$ are the areas of pillar surface and bottom surface, respectively, and $s_x$ and $n_x$ are the streamwise components of the unit vectors in tangential and normal directions, respectively. The equivalent shear stresses on the structured wall of these forces is obtained by dividing the forces by the horizontal area of one structure cell ($A_c$). The equivalent shear stresses can be further normalized by the shear stress at smooth bottom surface ($\tau_s = \mu U_0/H$),

$$\tilde{T} = \frac{F}{A_c(\mu U_0/H)} \tag{15}$$

Therefore, the equivalent shear stress of total drag in the normalized form is written as,

$$\tilde{T}_{tot} = \tilde{T}_{p,s} + \tilde{T}_{p,p} + \tilde{T}_{b,s} \tag{16}$$

where $\tilde{T}_{tot}$, $\tilde{T}_{p,s}$, $\tilde{T}_{p,p}$ and $\tilde{T}_{b,s}$ are the normalized equivalent shear stresses of $\tilde{F}_{tot}$, $\tilde{F}_{p,s}$, $\tilde{F}_{p,p}$, and $\tilde{F}_{b,s}$. The three components ($\tilde{T}_{p,s}$, $\tilde{T}_{p,p}$ and $\tilde{T}_{b,s}$) couple with each other, and each depends on both Reynolds number and surface topologies. Here only the Reynolds number effect is considered.

Figure 14 shows the dependence of $\tilde{T}_{tot}$, $\tilde{T}_{p,s}$, $\tilde{T}_{p,p}$ and $\tilde{T}_{b,s}$ on $Re$ in the range from $Re = 0.33$ to $100$. The symbols indicate the simulations conducted in this study. The equivalent shear stress of total drag ($\tilde{T}_{tot}$) which is determined by three components, remains roughly at a constant level in the range of $Re$ considered in this study. Due to the blocking of pillar array, the equivalent shear stress at the bottom surface ($\tilde{T}_{b,s}$) is much smaller than the other two components. Due to the aforementioned influence of fluid inertia on the momentum transport in vertical direction, $\tilde{T}_{b,s}$ decreases with the increase in $Re$ when fluid inertia takes effect at a critical Reynolds number $Re_c$ ($20 < Re_c < 30$), as shown in Fig. 14(b). This is consistent with the analysis of $\partial \tilde{u}_x / \partial \tilde{z}$ in Fig. 11. When $Re$ is less than $Re_c$, the fluid inertia is weak, and $\tilde{T}_{tot}$, $\tilde{T}_{p,s}$, $\tilde{T}_{p,p}$ and $\tilde{T}_{b,s}$ remain at their respective constants. This $Re$ region can be considered as a Stokes flow region, where the influence of fluid inertia on these equivalent shear stresses is not obvious. When $Re$ is greater than $Re_c$, the fluid inertia takes effect, which makes the overhead flow tilts downward and hit the pillar tips more strongly, as shown in Fig. 7 and 8. As a result, the component of pressure force of micro pillars ($\tilde{T}_{p,p}$) increases with the increase in $Re$. For the component of flow shear force of micro pillars ($\tilde{T}_{p,s}$), the curve shows that $\tilde{T}_{p,s}$ decreases with the increase in $Re$ and asymptotically approaches a constant value. In the nondimensional form, $\tilde{T}_{p,s}$ quantifies the flow shear rate on pillar surface. The curve variation suggests that the equivalent flow shear rate is smaller for larger Reynolds number when fluid inertia takes effect. This observation seems to contradict the common understanding of the effect of fluid inertia which suggests that the larger the Reynolds number, the greater the flow shear rate.

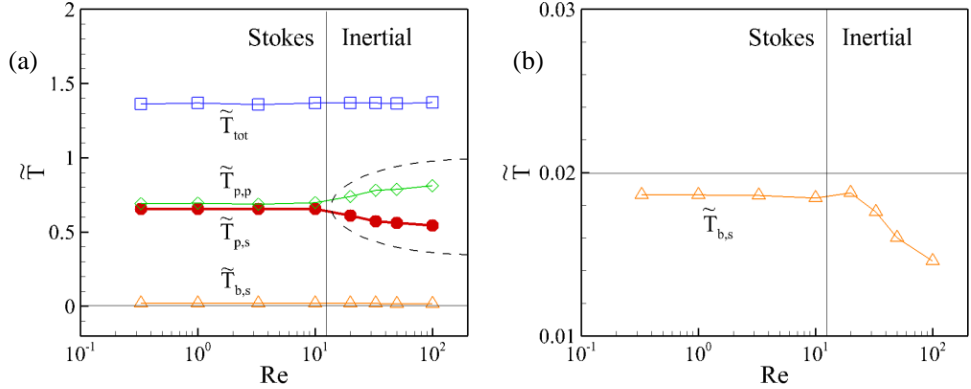

FIG. 14. Dependence of equivalent shear stress of total friction ($\tilde{T}_{tot}$), equivalent shear stresses of shear stress and pressure at pillar surface ($\tilde{T}_{p,s}$ and $\tilde{T}_{p,p}$), and shear stress at bottom plane ($\tilde{T}_{b,s}$), on Reynolds number ($Re$). The right panel shows the dependence of $\tilde{T}_{b,s}$ on $Re$.

To further investigate this seemingly abnormal trend of shear stress with respective to $Re$, we present in Figure 15 the iso-contours of vertical velocity ($\tilde{u}_z$) on a streamwise plane through pillar centerlines, and the spanwise velocity ($\tilde{u}_y$) on a horizontal plane at $\tilde{z} = 0.9$, for $Re = 1, 33$ and $100$, respectively. As shown in Fig. 15(a), the contours of $\tilde{u}_z$ exhibit a good anti-symmetry between the windward and leeward sides of the pillar at $Re = 1$, as a result of the small fluid inertia at low Reynolds numbers. For $Re = 33$, the anti-symmetrical pattern is broken. Around the pillar tip, the contour lines shifts downstream, and the magnitude of vertical velocity becomes smaller on the leeward side. This suggests that the fluid flow does not return to the same vertical level as it climbs over the pillar tip, because the fluid inertia tends to maintain the fluid motion along a straight line and prevent the fluid from flowing along the curved pillar tip. This motion pattern reduces the magnitude of flow shear rate on the leeward surface of pillar tip, and further reduces the equivalent shear stress of flow shear at pillar surface ($\tilde{T}_{p,s}$). Below the pillar tip, the magnitude of $\tilde{u}_z$ on the windward side at $Re = 33$ is larger than that at $Re = 1$, due to the downward tilt of the overhead flow (Fig. 7, 8 and 10). On the leeward side, the change in $\tilde{u}_z$ is not obvious. When $Re$ increases to 100, the asymmetry is more obvious, and the magnitude of $\tilde{u}_z$ further decreases on the leeward side of the pillar tip, which further decrease the magnitude of flow shear rate. It should be noted that the effect of fluid inertia on the flow on the surface of pillar tip and on the flow above the gaps between streamwise neighboring pillars is different. Above the gaps, the fluid inertia tends to maintain the downward motility of the overhead flow after climbing over the pillar tips and creates an overall downward tilt of the streamlines (Fig. 8). Along the surface of hemispherical pillar tips, the fluid inertia tends to maintain the flow straight-line motion, and prevents the fluid from flowing along the curved pillar surface.

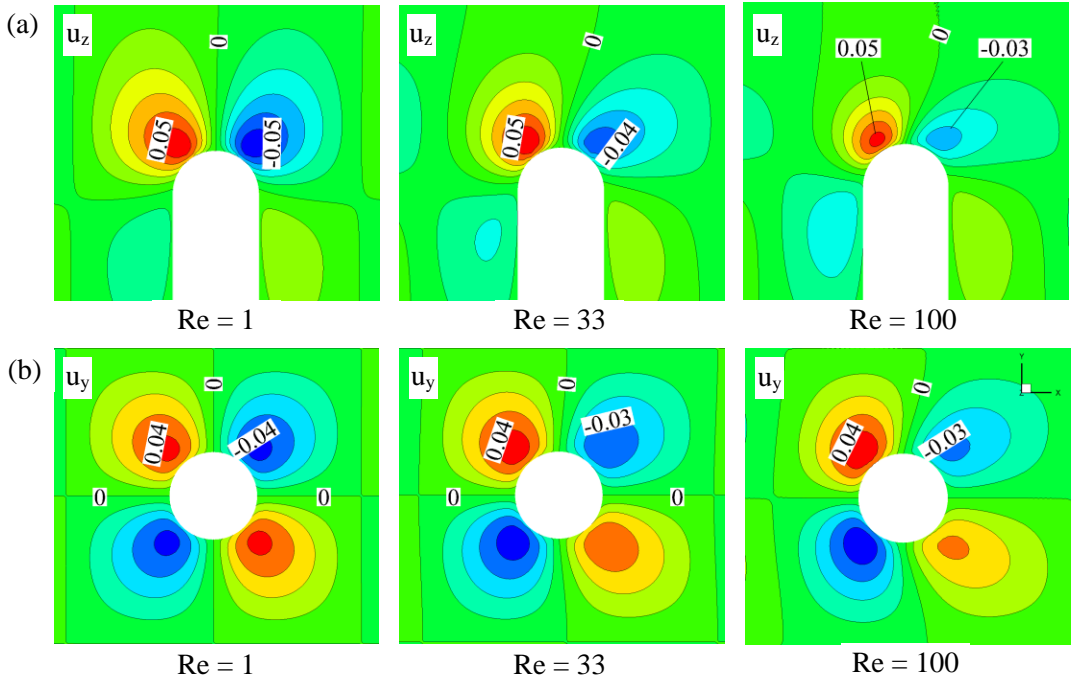

FIG. 15. Flow velocity components around pillar tip for $Re = 1$, 33 and 100. (a) vertical velocity $(\tilde{u}_z)$ at a streamwise plane through pillar centerlines, and (b) spanwise velocity $(\tilde{u}_y)$ at a horizontal plane at $\tilde{z} = 0.9$.

Figure 15(b) shows the iso-contours of spanwise velocity $(\tilde{u}_y)$ on a horizontal plane at $\tilde{z} = 0.9$. Similar to $\tilde{u}_z$ at the vertical plane shown in Fig. 15(a), the pattern of $\tilde{u}_y$ exhibits a good anti-symmetry between the windward and leeward sides of the pillars at $Re = 1$. At $Re = 33$, the anti-symmetrical pattern is broken, and the magnitude of $\tilde{u}_y$ is smaller on the leeward side due to the fluid inertia. The decrease in the magnitude of $\tilde{u}_y$ causes the decrease in the flow shear rate on the leeward surface, and further reduces the equivalent shear stress of flow shear on pillar surface $(\tilde{T}_{p,s})$. When $Re$ increases to 100, this phenomenon is more prominent, that is, the magnitude of $\tilde{u}_y$ further decreases on the leeward side. Due to the downward movement of the overhead flow, the area of larger $\tilde{u}_y$ on the windward side increases as $Re$ increases from 1 to 100, which implies that the magnitudes of $\tilde{u}_y$ and flow shear rate on the windward side increase with the increasing $Re$. However, this increase in flow shear rate is still not enough offset the weakening effect of the decreasing flow shear rate on the leeward side on $\tilde{T}_{p,s}$. Therefore, $\tilde{T}_{p,s}$ decreases with the increasing $Re$, as shown in Fig. 14.

Fig. 16 shows the meridional velocity $(\tilde{u}_\theta)$ along the 45° line on the leeward side at the streamwise plane and the azimuthal velocity $(\tilde{u}_\phi)$ along the 45° line on the leeward side at the horizontal plane at $\tilde{z} = 0.9$. As shown in the figure, the curve slope becomes smaller as the Reynolds number increases from 1 to 100 for both $\tilde{u}_\theta$ and $\tilde{u}_\phi$. This confirms the conclusion that the flow shear rate at the pillar surface on the leeward side decreases with the increase in Reynolds number.

## V. Conclusions

Through high-fidelity numerical simulation based on the lattice Boltzmann method, we have conducted an in-depth study on the simple shear flow over regularly arranged micro pillars. As the first part of a series of studies, this paper focuses on several fundamental issues essential to the understanding of the transport mechanism of the fluid flows over micro-structured surfaces: (1) characteristics of a simple shear flow over quadrilateral array of micro pillars, (2) effect of fluid inertia on the basic flow pattern, and (3) decomposition of the complex surface friction.

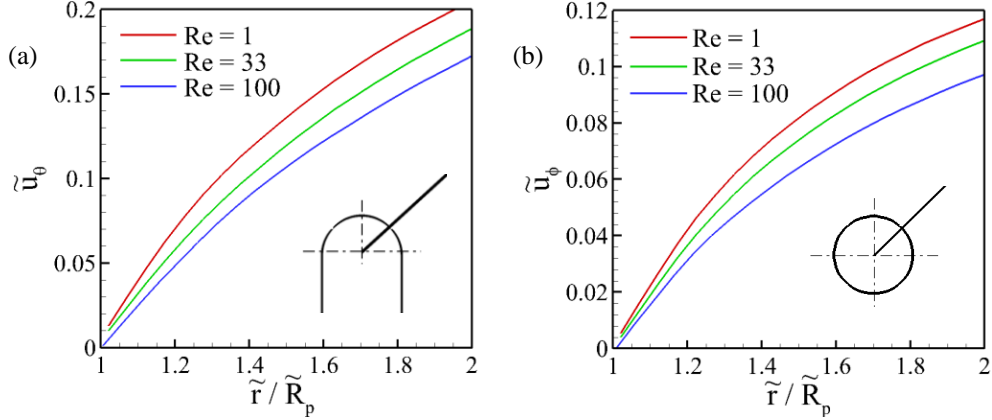

FIG. 16. Distributions of (a) meridional velocity along the 45° line on the leeward side at the streamwise plane through pillar centerlines, and (b) azimuthal velocity along the 45° line on the leeward size at the horizontal plane at $\tilde{z} = 0.9$.

As the first focus of this paper, the basic characteristics of a simple shear flow over quadrilateral array of micro pillars were explored in depth. It was found that the streamwise velocity changes linearly with the vertical coordinate in the region above the pillar tips. Below the pillar tips, the fluid velocity is significantly reduced and a microscale eddy is generated in the gap between each pair of streamwise neighboring pillars. The flow recirculation of the micro eddies and the flow oscillation of the overhead flow climbing over the pillar tips create a local advection, providing a mechanism of momentum, heat and mass transport in the wall-normal direction. The results show that the fluid inertia plays an important role in determining the flow patterns. At smaller Reynolds number, the fluid inertia is weak and the flow patterns on the windward and leeward sides of micro pillars are symmetrical about the pillar center. When the Reynolds number is sufficiently large, the fluid inertia takes effect and breaks the symmetrical pattern. The fluid inertia makes the overhead flow tilt downward, forming a spiral long-range advection between the fluid flow above pillar array and the flow among micro pillars. The long-range advection not only changes the characteristics of flow evolution but also changes the momentum transport from the upper fluid to the fluid among micro pillars and to the micro pillars. On the micro-structured walls, the total friction includes the reaction force of micro pillars due to flow shear stress at pillar surfaces, the reaction force of micro pillars due to flow pressure at pillar surfaces, and the reaction force of bottom plane due to flow shear on bottom surface. For larger Reynolds numbers, fluid inertia prevents the fluid from flowing along the curved surface of micro pillars and reduces the equivalent shear stress of the reaction force due to flow shear on pillar surface. At the same time, the fluid inertia makes the overhead flow impact the windward side of micro pillars more strongly and therefore increases the equivalent shear stress of the reaction force due to flow pressure on pillar surfaces.

The phenomena and mechanisms discovered in this work are of groundbreaking significance for the research and applications involving fluid flows over micro-structure surfaces, and provide critical guidelines for the precise control of flow behaviors and surface friction.